**On the Chemistry and Physical Properties of Flux and Floating Zone Grown SmB$_6$ Single Crystals**

W. A. Phelan,[1,2,3*] S. M. Koohpayeh,[2] P. Cottingham,[1,2,3] J. A. Tutmaher,[1,2,3] J. C. Leiner,[4] M. D. Lumsden,[4] C. M. Lavelle,[5] X. P. Wang,[6] C. Hoffmann,[6] M. A. Siegler,[1] and T. M. McQueen[1,2,3*]

[1]Department of Chemistry, Johns Hopkins University, Baltimore, MD 21218, USA
[2]Institute for Quantum Matter, Department of Physics and Astronomy, Johns Hopkins University, Baltimore, MD 21218, USA
[3]Department of Materials Science and Engineering, Johns Hopkins University, Baltimore, MD 21218, USA
[4]Quantum Condensed Matter Division, Oak Ridge National Laboratory, Oak Ridge, TN 37831, USA
[5]Applied Nuclear Physics Group, Johns Hopkins University, Applied Physics Laboratory, Laurel, MD, 20723, USA
[6]Chemical and Engineering Materials Division, Neutron Sciences Directorate, Oak Ridge National Laboratory, Oak Ridge, TN 37831, USA
[*]wphelan2@pha.jhu.edu and mcqueen@jhu.edu

**Abstract**

Recent theoretical and experimental findings suggest that the long-known but not well understood low temperature resistance plateau of SmB$_6$ may originate from protected surface states arising from a topologically non-trivial bulk band structure having strong Kondo hybridization. Yet other studies have ascribed this feature to impurity phases, sample vacancies, and surface reconstructions. Given the typical methods used to prepare SmB$_6$ single crystals, the flux and floating zone procedures, these ascriptions should not be taken lightly. Here, we demonstrate how compositional variations and observable amounts of impurity phases in SmB$_6$ crystals grown by floating zone and flux affect the physical properties. From neutron diffraction and X-ray computed tomography experiments, we observe that a $^{154}$Sm$^{11}$B$_6$ crystal prepared using aluminum flux contains co-crystallized, epitaxial aluminum. A large, nearly stoichiometric crystal of SmB$_6$ was successfully grown using the float-zone technique; upon continuing the zone melting, samarium vacancies are introduced. These samarium vacancies drastically alter the resistance and plateauing magnitude of the low temperature resistance compared to stoichiometric SmB$_6$. These results highlight that small presences of impurity phases and compositional variations must be considered when collecting and analyzing physical property data of SmB$_6$. Finally, a more accurate samarium-154 coherent neutron scattering length value, 8.9(1) fm, is reported.

**Introduction**

The lanthanide hexaborides, $Ln$B$_6$ ($Ln$ = La-Nd, Sm, and Eu-Ho, Yb, and Y), comprise a fascinating class of materials which crystallize with a primitive cubic structure ($a \sim 4.1$ Å, $V \sim 70$ Å$^3$, $Z = 1$, and SG = $Pm$-$3m$).[1] In the case of $Ln$ = Sm, many theoretical and experimental studies of SmB$_6$ have been performed for more than four decades and have focused on its samarium ion mixed valence nature (Sm$^{+2}$ and Sm$^{+3}$), Kondo insulating behavior, and the mysterious plateau commonly observed in the low temperature regime of many temperature dependent resistance datasets.[2,3] More recently, theoretical and experimental studies have focused on the possibility that Kondo insulating SmB$_6$ harbors non-trivial topological protected surface states, which might provide an explanation for the low temperature resistance plateau.[4-23] However, previous literature claims that this remnant metallic feature results from combinations of impurity phases and compositional variations.[24,25]

To determine the possible contributions of impurity phases and compositional variations to the low temperature resistance behavior, a central need is high quality single crystals of SmB$_6$ with precisely defined and controlled stoichiometry. The main procedures used to synthesize single crystals of SmB$_6$ are the floating zone ($FZ$) and flux growth ($FG$) methods. In general, the $FG$ technique is used to grow crystals below their melting temperature, and as a result, high temperature decomposition can be prevented. However, the major disadvantages of this technique can include crystal products that are small in size, the presence of flux inclusions within the crystal, contamination from the melt container, and possible inhomogeneities in the crystal due to inconsistent linear growth rates of different facets that develop during growth. In contrast to the $FG$ technique, the use of the crucible-and flux-less $FZ$ technique circumvents the problem of the incorporation of impurities from the crucible and flux materials; further, the grown crystals are quite large and can be manipulated easily. However, preparing single crystals $via$ this technique, which is widely used for the crystal growth of the congruently or near congruently melting compounds, can be more complicated for the materials with instabilities near their melting temperature.[26]

Here, we report the results of a comparative study of *FZ* and *FG* single crystals of $SmB_6$. We find large systematic changes in the occurrence of the low-temperature resistance plateau with small but systematic changes in the lattice parameters of *FZ* prepared $SmB_6$ along the length of a large single crystal, showing that small changes in composition have large effects on this and other physical properties of $SmB_6$. Further, single crystal neutron diffraction studies of a doubly isotope enriched *FG* $^{154}Sm^{11}B_6$ single crystal revealed the presence of epitaxially oriented, co-crystallized aluminum. These results show that metallic aluminum incorporated into *FG* crystals of $SmB_6$ will have an effect on the physical properties of crystals prepared using this method. Finally, using the neutron diffraction experiments, we have determined a more accurate coherent neutron scattering length value of 8.9(1) fm for samarium-154.

**Results and Discussion**

**Chemistry and Physical Properties.** As mentioned above, the *FG* procedure is a common method used to grow $LnB_6$ single crystals. A typical photograph $SmB_6$ crystal grown using the *FG* procedure is shown in Figure 1a. The crystal has several facets and is of the appropriate size for diffraction experiments and physical properties measurements. Under magnification, portions of surface of this crystal are coated with a lustrous, metallic substance (likely residual aluminum flux), which is typically etched away prior to physical property measurements.

Using single crystal neutron diffraction, which effectively probes the entire sample volume due to the penetrating power of neutrons, we find the residual aluminum is not confined to the surfaces of this doubly isotope enriched crystal, which was grown using samarium-154 and boron-11 (The least neutron absorbing, stable isotopes of samarium and boron.). Figure 2a shows a neutron precession image along the (00*l*) and (*h*00) directions. A set of companion peaks (white arrow) accompany the main $SmB_6$ Bragg reflections. To identify this satellite crystallite, a powder-like neutron diffraction histogram was obtained from the radial integration of the single crystal neutron diffraction data, Figure 2b. The asterisk denotes the reflections from the secondary phase. We identify it as aluminum metal for three reasons: first, the lattice parameter ($a$ = 4.053(3) Å) is within 0.09-0.3% of the literature values for aluminum; second, the systematic absences are consistent with the *Fm*-3*m* space group of Al; and third, there are no other known compounds that can explain these diffraction peaks and the observed lattice constant containing some combination of *Ln*, B, C, V, and Al (*Ln* = La-Nd and Sm-Lu). An aluminum/SmB$_6$ co-Rietveld refinement to the powder averaged data showed the ratio of aluminum:SmB$_6$ was ~ 4 wt%:96 wt% or ~ 25 mol%:75 mol%. Such a significant amount of well-ordered aluminum cannot originate solely from a surface coating, and instead demonstrates that aluminum inclusions are present within this SmB$_6$ *FG* crystal. This is not surprising given the similar crystal structures of SmB$_6$ and aluminum (cubic crystal symmetry with the *a*-lattice parameters for both being ~ 4.1 Å), the large excess of aluminum flux required for growths given the solubility of boron under typical *FG* conditions, and the previous reports of metallic inclusions during *FG* crystal polishing.[24,27]

To check the validity of the neutron experiments and data analysis, X-ray computed tomography (CT) volumetric data of the *FG* $^{154}$Sm$^{11}$B$_6$ crystal were obtained using this crystal. A resulting image of the 3D reconstruction is shown above the precession image in Figure 2a, and a movie generated from the reconstruction of the frames has been added to the Supplementary Information (Figure S1.). Regions composed of materials with high Z (atomic number) and/or high number density will attenuate the X-ray beam more strongly than lower Z, low density regions. Clearly, this crystal is comprised of two materials: one with a high average Z (atomic number, dark contrast), and a second with low, but non-negligible, average Z (light contrast). This is exactly what would be expected for a SmB$_6$ crystal with Al inclusions: the SmB$_6$ corresponds to the high Z regions, and the Al the low Z regions. Several features of the aluminum inclusions are evident: they are not confined to the surface, but instead are included deep within the crystal, have well-defined facets oriented non-randomly with respect to the host, and are present in an amount consistent with aluminum/SmB$_6$ co-Rietveld refinements to the neutron data.

A surprising finding from our neutron and X-ray CT data is that the aluminum inclusions are not randomly oriented, but instead maintain a nearly epitaxial registry with the hexaboride, and have non-negligible size. These inclusions, which will not necessarily affect the electrical transport at low temperature (plateau region) as they are embedded in a host insulating material, would provide an alternate explanation for the origin of high frequency and light mass quantum oscillations observed in torque magnetometry experiments on *FG* $SmB_6$ samples. However, given their size, variability in the number of such inclusions is likely and should be checked on a crystal by crystal basis.

In order to avoid flux inclusions and grow a larger, more homogeneous and higher purity crystal of $SmB_6$, which is reported to be a congruently melting compound, the *FZ* technique was successfully utilized to produce the $SmB_6$ single crystal shown in Figure 1b. This crystal, measuring roughly 8 cm in length, is considerably larger than $SmB_6$ single crystals that can be prepared using *FG*. Similar to *FG* $SmB_6$, there are also composition variations in the *FZ* crystal.

To understand the origin of the composition variation, four individual cuts were taken from the *FZ* grown crystal (Figure 3b Inset), and a portion of each individual cut was used for synchrotron powder X-ray diffraction while another portion was used for physical property measurements followed by trace elemental analysis. The resulting synchrotron powder X-ray diffraction data and fit for cut 1 is shown in Figure 3a; similar quality fits are obtained for cuts 2-4 (Figure S2 in the Supplementary Information). Final parameters for all refinements are given in Table 1. There is a systematic change in lattice parameters across the cuts. Going from cut 1 to cut 3, the lattice parameters decrease, while the lattice parameter for cut 4 is similar to cut 3. This is shown graphically in Figure 3b and Figure S4. The refined *a*-lattice parameter value for the polycrystalline feed rod, as confirmed by powder X-ray diffraction, is shown as a blue dashed line, which lies between the start (cut 1) and end (cut 4) of the grown crystal. This is consistent with the temperature-composition phase diagram for the Sm-B system which indicates that by fully melting a feed rod with the composition either near to the stoichiometric ratio (Sm/B : 1/6) or over the range of Sm to B atomic percent ratios from ~14.3% Sm-85.7% B (Sm/B : 1/6) to ~9.0% Sm-91.0% B (Sm/B : 1/10), the first solidified crystal would have the highest possible Sm to B ratio within the $SmB_6$ crystal structure (at ~14.3 at%Sm-85.7at%B).[28] This explains how a larger lattice parameter of 4.1343 Å obtained for the initially growing crystal occurs with the use of a polycrystalline feed rod with a slightly smaller lattice parameter of 4.1333 Å.

Attempts to directly identify the precise origin of the compositional change from Rietveld refinements or trace elemental analysis were unsuccessful. When the boron and samarium occupancy parameters for all cuts were allowed to individually float during separate Rietveld refinement cycles, no significant deviation from unity was noticed for either on any dataset. As such, for all final refinements the occupancy parameters for boron and samarium were fixed at unity. Trace elemental analysis via glow discharge mass spectrometry (GDMS) also showed no systematic trends in total, rare earth, or transition metal impurity levels as a function of rod length, Figure S6 and Table S4.

Nonetheless, the lattice parameters listed in Table 1, combined with previous robust elemental analyses, do provide insight into the composition differences between the different *FZ* cuts. The lattice parameter for cut 1 is in excellent agreement with the lattice parameter ($a$ = 4.1342(5) Å) determined for stoichiometric $SmB_6$ as reported by Tarascon *et. al.*,[29] and close to Paderno *et. al.*'s most stoichiometric sample of $SmB_6$ (a = 4.13334(2) Å).[30] The reduction of the lattice parameter in subsequent cuts is consistent with the formation of samarium vacancies, based on Paderno *et. al.*'s careful chemical analyses which found that the *a*-lattice parameter decreases as the samarium content decreases. While the variation between cuts 1 to 4 is small, less than 1% or $Sm_{1-x}B_6$ (x = 0.01), based on the *a*-lattice parameter of 4.1317(2) Å for $Sm_{0.97}B_6$ and assuming Vegard's law applies, we have evidence that this compositional variation is due to vaporization during growth. During each growth, we observe vaporization of a small (~1%) amount of the rod material. Using laboratory powder X-ray diffraction, the vaporized powder was confirmed to be a multiphase mixture of $SmB_6$ and $SmB_4$, i.e. samarium rich compared to the feed rod, implying a relative loss of samarium in the growing crystal (Figure S3 in the Supplementary Information). While this loss of material could be due to stoichiometric $SmB_6$ being an incongruent melter, a more likely explanation is a slight non-stoichiometry in the feed rods themselves: the lattice parameter for the polycrystalline feed rod starting material is slightly nonstoichiometric relative to the most stoichiometric portion of our grown *FZ* single crystal (cut 1). These results, however, are in contrast to a recent $SmB_6$ thin-film study, where the authors found that the lattice parameter increased with increasing samarium vacancy content. This result is consistent with lattice mismatch effects on the MgO substrate, where the higher samarium vacancy nanocrystal films become strained, and as more samarium is added the lattice mismatch is reduced, and thus the size of the lattice constant decreases.[31]

The structural parameters can also be used to independently determine the samarium valence. The structure of $SmB_6$, shown in Figure 4a, adopts the $CaB_6$ structure-type, or more descriptively is analogous to CsCl where the Sm atoms occupy the vertices of a cube and the $B_6$ octahedra lie in the center of the unit cell. An additional structural feature shown in Figure 4a is the bonding between $B_6$ octahedra (inter), which is shorter in distance compared to the boron-boron bonds forming the $B_6$ octahedra (intra), that forms a three dimensional interconnected cage structure. The intra-octahedral boron-boron bond distances normalized by the inter-octahedral boron distances (Intra B-B/Inter B-B) for selected hexaborides versus different charges of the $B_6$ cluster ($B_6^x$; x = -4, -3, and -2) are plotted in Figure 4b. Also shown are the Intra B-B/Inter B-B values for all $SmB_6$ *FZ* cuts and *FG* $SmB_6$ versus charge of the $B_6$ cluster for all charges (dashed line). Relative to the -4 ($ThB_6$)[32] and -2 ($CaB_6$, $SrB_6$, $BaB_6$, and $EuB_6$)[33-35] charges, the charge on the $B_6$ cluster for all $SmB_6$ samples appears to most closely resemble -3 ($NdB_6$ and $LaB_6$).[35,36] However, the dashed line falls slightly below the Intra B-B/Inter B-B values for $NdB_6$ and $LaB_6$,[34,36] and this is consistent with a partial mixed valency. Applying a lever-type rule, an estimate of the $Sm^{+3}$:$Sm^{+2}$ ratio for all samples was determined to be ~ 80%:20%. Given that the Intra B-B/Inter B-B values for all $SmB_6$ samples lie on the dashed line in Figure 4b, this suggests that while the composition changes along the *FZ* crystal have noticeable effects on the lattice parameters, the overall samarium vacancies and valence are not drastically altered along the length of the crystal (this is not unexpected as the samarium content is changing by <1%). It is interesting to note that these data, which are consistent with a $Sm^{+2.80}B_6$ scenario, are not necessarily in quantitative agreement with the findings of other experiments like X-ray absorption spectroscopy (XAS) where those data and analysis show that bulk $SmB_6$ has a mixed valence nature closer to $Sm^{+2.60}B_6$.[29,37] The most plausible explanation for this difference, if real, is that in addition to formal changes in the oxidation states, there are configurational changes that occur on samarium without a change of formal oxidation state.

To see how the changes in composition along the *FZ* crystal affect the physical properties of SmB$_6$, temperature dependent resistance data were collected for each *FZ* cut. The 300 K normalized resistance data for all cuts are plotted from *T* = 2 – 300 K in Figure 5. At higher temperatures the data from all cuts closely resemble the resistance data reported for a majority of SmB$_6$ samples. That is, the resistances are roughly temperature independent from 300 K to 100 K, increase dramatically below 40 K, and have some degree of plateauing below ~10 K. We find systematic variations in the low temperature behavior with changing samarium vacancy concentration. The low temperature remnant metallicity decreases from cut 1 to cut 3 along with the overall normalized resistance values, while the features for cuts 3 and 4 are seemingly identical. The degree of reproducibility of these resistance curves is highlighted in Figure S5. This figure shows a second set of measurements where new contacts replaced the old contacts of the original cuts 1-3 (open colored circles) and resistances were measured using a new cut between the location of the original cut1 and cut 2 and a new cut beyond the location of the original cut 4 (filled gray squares). We have previously shown that electron doping SmB$_6$ *via* the replacement of carbon for boron in the boron sub-lattice can induce a low temperature resistance plateau.[22] Samarium vacancies will introduce holes, rather than electrons, and from our previously proposed density of states model, this is expected to reduce the degree of resistance plateauing in SmB$_6$.[22] Taken together, these observations suggest that the composition along the *FZ* crystals is being changed systematically, and that these small changes in compositions produce noticeable changes in physical properties. These results and our previous SmB$_6$ dopant study highlight the utility of the *FZ* technique to systematically control crystal stoichiometry in SmB$_6$.[22]

**$^{154}$Sm Neutron Scattering Length Determination.** The previously best determined coherent neutron scattering length value (8.0(1.0) fm) for samarium-154 ($b_{154_{Sm}}$) has a large uncertainty that makes it unsuitable for detailed refinements of single crystal neutron diffraction data, and thus for determination of any Sm:B non-stoichiometry.[38] We thus performed the necessary experiments to provide a refined number with an order of magnitude reduced uncertainty.

Two small crystal pieces were removed from the doubly isotope enriched $^{154}Sm^{11}B_6$ *FG* single crystal shown in Figure 1a. The first piece was used to determine the various samarium and boron isotope ratios *via* ICP-MS, while the second piece was used to collect a large and highly redundant single crystal X-ray diffraction dataset (see SI). Refinements to this dataset returned the best goodness-of-fit statistics when the composition of this crystal was set to $SmB_{5.88}$ and all other parameters were allowed to float (Table S1 and S2). Using this composition, the known coherent scattering lengths for 10-boron ($^{10}B$) and 11-boron ($^{11}B$),[38] and determining the overall coherent neutron scattering length for boron in this crystal *via* $b_B = w_1{}^{10}B + w_1{}^{11}B = 6.4(1)$ fm where $w_1$ and $w_2$ were determined from the ICP-MS data; an overall coherent neutron scattering length of 8.85 fm was determined for the combination of samarium isotopes ($b_{x_{Sm}}$) in the doubly isotope enriched single crystal through a series of refinements to the $T = 295$ K neutron diffraction data. When the errors based on the X-ray refinements and compositions for $b_{x_{Sm}}$ and $b_B$ were taken into account, respectively, the overall coherent scattering length value is found to equal 8.9(1) fm. Finally, after setting this value equal to the weighted values (Again, these values were determined from the ICP-MS data.) of the known coherent scattering lengths of the samarium isotopes present in this crystal ($^{144}Sm$, $^{147}Sm$, $^{148}Sm$, $^{149}Sm$, $^{150}Sm$, and $^{152}Sm$),[38] a new value equal to 8.9(1) fm was determined for $^{154}Sm$ (Table S3).

Once the $b_B$ and $b_{x_{Sm}}$ were set to be 6.4(1) fm and 8.9(1) fm, respectively, refinements to the $T = 90$ K and 295 K neutron data were performed. The crystallographic parameters and refinement statistics for both temperatures are provided in Table 2, and the atomic fractional coordinates, site occupancies, and ADPs are given in Table 3. During separate refinements the samarium and boron occupancy parameters were allowed to float. In all refinements, the resulting occupancies were found to be within $3\sigma$ of unity, thus, the information in Tables 3 and 4 reflect refinements where the occupancies have been set to unity.

## Conclusions

Structural data for *FG* and *FZ* single crystals of $SmB_6$ have been compared. Observable amounts of impurity phases and compositional variations can have large effects on the physical properties of $SmB_6$ crystals prepared using *FG* and *FZ* procedures, respectively. In particular, epitaxially oriented aluminum inclusions are found in *FG* crystal and samarium vacancies found in *FZ* crystal. Further, the boron-boron bond distances determined from our X-ray and neutron refinements are consistent with mixed valency with a formula close to $Sm^{+2.80}B_6$. The changes in boron-boron bond distances for $SmB_6$ and other hexaborides determined using temperature dependent X-ray and neutron scattering data would be interesting to study in order to see if changes in the charge transfer to the $B_6$ cage for $SmB_6$ occur relative to other hexaborides. Finally, we have determined that the coherent neutron scattering length for samarium-154 equals 8.9(1) fm, a value an order of magnitude more accurate than previously known.

**Methods**

  **Synthesis**. A single crystal of $SmB_6$ with approximate dimensions of 80 mm in length and 6 mm diameter was prepared from polycrystalline rods of $SmB_6$ (Testbourne Ltd, 99.9%) using a four-mirror optical floating zone furnace (Crystal Systems Inc FZ-T-12000-X-VPO-PC) with 4 × 3 kW Xe lamps as the heating source. The crystal growth was performed by melting the polycrystalline feed rod onto a seed rod, then running the molten zone in an upward direction along the feed rod (defining the growth direction) at a zoning rate of 10 mm/h, under flowing ultra-high purity argon at a pressure of 2 bar with a flow rate of 2 L/min, and the rotation rate of 10 rpm for the growing crystal. Only one zone pass was required for the growth. Slices of the crystal were cut close to the [100] orientation using a diamond saw.

  The *FG* single crystal of doubly isotope enriched $SmB_6$ ($^{154}Sm^{11}B_6$), needed for the single-crystal neutron diffraction experiments as naturally occurring samarium and boron are largely comprised of isotopes that absorb neutrons, was provided by Oak Ridge National Laboratory (ORNL).

  **Powder X-ray Diffraction**. Slices cut from a large *FZ* grown $SmB_6$ single crystal were ground using a stainless steel mortar and pestle. A small amount of a powdered silicon standard (SG = *Fd*-3*m* and *a* = 5.43102 Å) was added to the resulting $SmB_6$ powder for each cut. To acquire a high intensity-high resolution powder X-ray diffraction dataset, synchrotron powder X-ray diffraction data were obtained at *T* = 295 K using the 11-BM beam line (λ = 0.4136820 Å) at the Advanced Photon Source within Argonne National Laboratory.[39] The data points were collected over a 2θ range 0.5º - 50º with a step size of 0.001° and step time of 0.1 seconds. Le Bail fits and Rietveld refinements were conducted using the GSAS/EXPGUI software to optimize the lattice/instrumental (GU, GV, and GW) and structure parameters for the $SmB_6$ models, respectively.[40,41] The crystallographic parameters and refinement statistics for all cuts are provided in Table 1.

All laboratory powder X-ray diffraction patterns were collected using Cu $K\alpha$ radiation on a Bruker D8 Focus diffractometer with a LynxEye detector. The same silicon standard described above (SG = $Fd$-$3m$ and $a$ = 5.43102 Å) was mixed in with powder from the polycrystalline rods of $SmB_6$ purchased from Testbourne Ltd. Rietveld refinements were performed in TOPAS (Bruker AXS) to determine a lattice parameter of $a$ = 4.133387(2) Å.

**Single Crystal X-ray Diffraction**. A small piece of a doubly isotope enriched crystal of $^{154}Sm^{11}B_6$ was mounted onto a fiber using epoxy. All reflection intensities were measured under ambient conditions using a SuperNova diffractometer (equipped with an Atlas detector) employing Mo $K\alpha$ radiation ($\lambda$ = 0.71073 Å) under the CrysAlisPro software package (version 1.171.36.28, Agilent Technologies, 2012). Unit cell indexing and data reductions were performed using the CrysAlisPro software. The generation of the initial models and structure refinements were performed using SIR97 and SHELXL-2013, respectively.[42,43] The selection of the $Pm$-$3m$ was based on the observed Laue symmetry and the systematic absences. After the refinement of the atomic positions, the collected data were corrected for absorption using an analytic correction.[44] During the final stages of refinements the atomic displacement parameters (ADPs) were refined as anisotropic and weighting schemes were applied. The crystallographic parameters and refinement statistics are provided in Table S1, and the atomic fractional coordinates, site occupancies, and ADPs are given in Table S2.

**Single Crystal Neutron Diffraction**. Single crystal neutron diffraction experiments were performed using the TOPAZ beam line at the Spallation Neutron Source at ORNL using a doubly isotope enriched crystal of $^{154}Sm^{11}B_6$. A $FG$ crystal with dimensions of 1.05 × 1.10 × 1.55 mm was mounted onto a Kapton covered vanadium post with Loctite instant adhesive (495) and positioned onto the goniometer. Data collections were performed at $T$ = 90 K and 295 K in wavelength-resolved time-of-flight (TOF)

Laue mode using neutrons with a wavelength range of $\lambda = 0.6\text{-}3.5$ Å. To ensure good coverage and redundancy for each data collection, data were collected with 13 detectors and using 13 to 15 crystal orientations, which were selected by evaluation with CrystalPlan software,[45] with collection times of approximately 3 hours per orientation. The integrated raw Bragg intensities were obtained using the 3-D ellipsoidal Q-space integration method in Mantid.[46] Data were corrected for background and detector efficiency. Data reduction including, Lorentz, neutron TOF spectrum, and absorption corrections was carried out with the local ANVRED2.[47] The reduced data were saved in SHELX HKLF2 format in which the wavelength is recorded separately for each individual reflection, and the reduced data were not merged as a consequence of the saved format. The crystallographic parameters and refinement statistics for both temperatures are provided in Table 2, and the atomic fractional coordinates, site occupancies, and ADPs are given in Table 3

**Mass Spectrometry**. Inductively coupled plasma mass spectrometry (ICP-MS) experiments and data analysis were performed at the Water Quality Center at Trent University in Ontario, Canada to determine the concentrations of the differing samarium and boron isotopes in the *FG* doubly isotope enriched $^{154}Sm^{11}B_6$ single crystal (See Table S3).

Glow discharge mass spectrometry (GDMS) and data analysis were performed by Evans Analytical Group to determine the concentrations of elements in the starting material, cut 1, cut 2, and cut 3 (See Table S4.).

**X-ray computed tomography**. The X-ray computed tomography (CT) data was collected on the $^{154}Sm^{11}B_6$ flux grown single crystal using a Bruker Skyscan 1172G. The source was set to 100kV/57 μA. Frames were collected in 0.5 degree steps using a 500 μm Al + Cu filter and SHT 11 Mp camera, with averaging of 100, 1.48 s exposures per angle and median filtering for 2.21 μm nominal resolution. Reconstruction was performed using the associated software. The final images were generated in attenuation mode, with contrast adjusted to visualize low Z inclusions.

**Physical Properties**. Temperature dependent resistance data were collected using the resistivity option of a 9-Tesla Quantum Design Physical Property Measurement System (PPMS). The measurements were performed using a standard four-probe method, where platinum leads were mounted in a linear configuration onto the crystals using silver epoxy. All crystals were bar shaped and had roughly the same geometric factors (length ~ 1.00 mm and cross-sectional area ~ 1.25 mm$^2$). For all measurements, very small excitation currents of 100 μA were used in order to avoid Joule heating effects.[22]

**Acknowledgments**

The work at the Institute for Quantum Matter (IQM) was supported by the U.S. Department of Energy, Office of Basic Energy Sciences, Division of Materials Sciences and Engineering under Grant No. DE-FG02-08ER46544. Use of the Advanced Photon Source at Argonne National Laboratory was supported by the U. S. Department of Energy, Office of Science, Office of Basic Energy Sciences, under Contract No. DE-AC02-06CH11357. The neutron diffraction data were collected at the Oak Ridge National Laboratory's Spallation Neutron Source; supported by the Division of Scientific User Facilities, Office of Basic Energy Sciences, U.S. Department of Energy, under contract DE-AC05 00OR22725 with UT Battelle, LLC. W.A.P would like to thank Collin L. Broholm for helping in the procurement of the flux grown $SmB_6$ crystal.

**Table 1.** Crystallographic parameters for the floating zone cuts of SmB$_6$ obtained from Rietveld refinements to the 11-BM data. The Sm and B atoms reside on the $1a$ (0,0,0) and $6f$ ($x$, ½, ½) Wyckoff position, respectively. The statistical uncertainties are given in parentheses.

|  | Cut 1 | Cut 2 | Cut 3 | Cut 4 |
|---|---|---|---|---|
| $T$ (K) | 295 | 295 | 295 | 295 |
| Space group | $Pm$-$3m$ | $Pm$-$3m$ | $Pm$-$3m$ | $Pm$-$3m$ |
| $a$ (Å) | 4.134309(2) | 4.133938(1) | 4.133288(1) | 4.133343(1) |
| $V$ (Å$^3$) | 70.67(1) | 70.65(1) | 70.61(1) | 70.62(1) |
| $Z$ | 1 | 1 | 1 | 1 |
| $x$ position (B) | 0.19985(18) | 0.19855(16) | 0.19962(15) | 0.19997(1) |
| $U_{Sm}$ (Å$^2$) | 0.007763(20) | 0.00802(2) | 0.007428(16) | 0.007360(12) |
| $U_B$ (Å$^2$) | 0.00258(13) | 0.00266(12) | 0.00177(11) | 0.00230(12) |
| [a]$R_p$ | 0.061 | 0.056 | 0.052 | 0.052 |
| [b]$R_{wp}$ | 0.080 | 0.073 | 0.067 | 0.072 |
| [c]$R_{exp}$ | 0.043 | 0.048 | 0.048 | 0.041 |
| [d]$\chi$ | 3.386 | 2.280 | 1.904 | 3.098 |

[a]$R_p = \sum |Y_o - Y_C| / \sum Y_o$; [b]$R_{wp} = [M/\sum w(Y_o^2)]^{1/2}$; [c]$R_{exp} = R_{wp}/(\chi^2)^{1/2}$; [d]$\chi = (M/N_{obs} - N_{va})^{1/2}$

**Table 2**. Crystallographic parameters for the flux grown $^{154}$Sm$^{11}$B$_6$ crystal obtained from model fits to the TOPAZ neutron data. The statistical uncertainties are given in parentheses.

| | | |
|---|---|---|
| Temperature (K) | 90 | 295 |
| Space group | *Pm-3m* | *Pm-3m* |
| *a* (Å) | 4.1306(2) | 4.1319(2) |
| *V* (Å$^3$) | 70.48(1) | 70.54(1) |
| *Z* | 1 | 1 |
| Collected Reflections | 977 | 1149 |
| Crystal Size (mm$^3$) | 1.05 × 1.10 × 1.55 | 1.05 × 1.10 × 1.55 |
| *GooF* | 1.30 | 1.13 |
| $R_1[F^2 > 2\sigma(F^2)]^a$ | 0.060 | 0.067 |
| $wR_2(F^2)^b$ | 0.166 | 0.190 |
| $\Delta\rho_{max}$ (fm Å$^{-3}$) | 0.99 | 1.17 |
| $\Delta\rho_{min}$ (fm Å$^{-3}$) | -2.59 | -3.21 |

$^a R_1(F) = \sum ||F_o| - |F_c||/\sum |F_o|$; $^b wR_2(F^2) = [\sum [w(F_o^2 - F_c^2)^2]/\sum [w(F_o^2)^2]]^{1/2}$

**Table 3.** Atomic fractional coordinates, site occupancies, and ADPs for the flux grown $^{154}$Sm$^{11}$B$_6$ crystal obtained from model fits to the TOPAZ neutron data. The statistical uncertainties are given in parentheses.

$T$ = 90 K

| atom | Wyckoff Site | x | y | z | Occupancy | $U_{11}$ (Å$^2$) | $U_{22}$ (Å$^2$) | $U_{33}$ (Å$^2$) |
|---|---|---|---|---|---|---|---|---|
| Sm1 | 1a | 0 | 0 | 0 | 1 | 0.00254(18) | 0.00254(18) | 0.00254(18) |
| B1 | 6f | 0.19984(7) | ½ | ½ | 1 | 0.00230(18) | 0.00361(18) | 0.00361(18) |

$T$ = 295 K

| atom | Wyckoff Site | x | y | z | Occupancy | $U_{11}$ (Å$^2$) | $U_{22}$ (Å$^2$) | $U_{33}$ (Å$^2$) |
|---|---|---|---|---|---|---|---|---|
| Sm1 | 1a | 0 | 0 | 0 | 1 | 0.0066(2) | 0.0066(2) | 0.0066(2) |
| B1 | 6f | 0.19964(6) | ½ | ½ | 1 | 0.00267(19) | 0.0044(2) | 0.0044(2) |

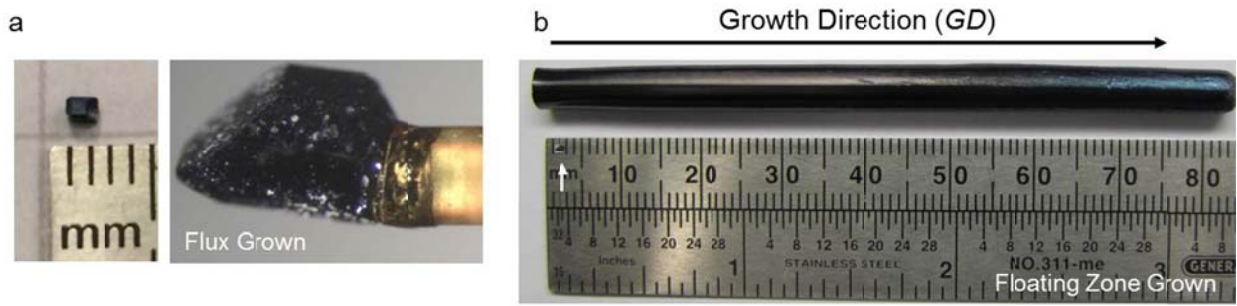

**Figure 1**. a) Pictures of a flux grown single crystal (right) and as mounted onto a Kapton covered vanadium post used for neutron diffraction measurements. b**)** A single crystal prepared *via* the floating zone procedure. Also shown is a scaled version of the picture of the flux grown single crystal (white arrow).

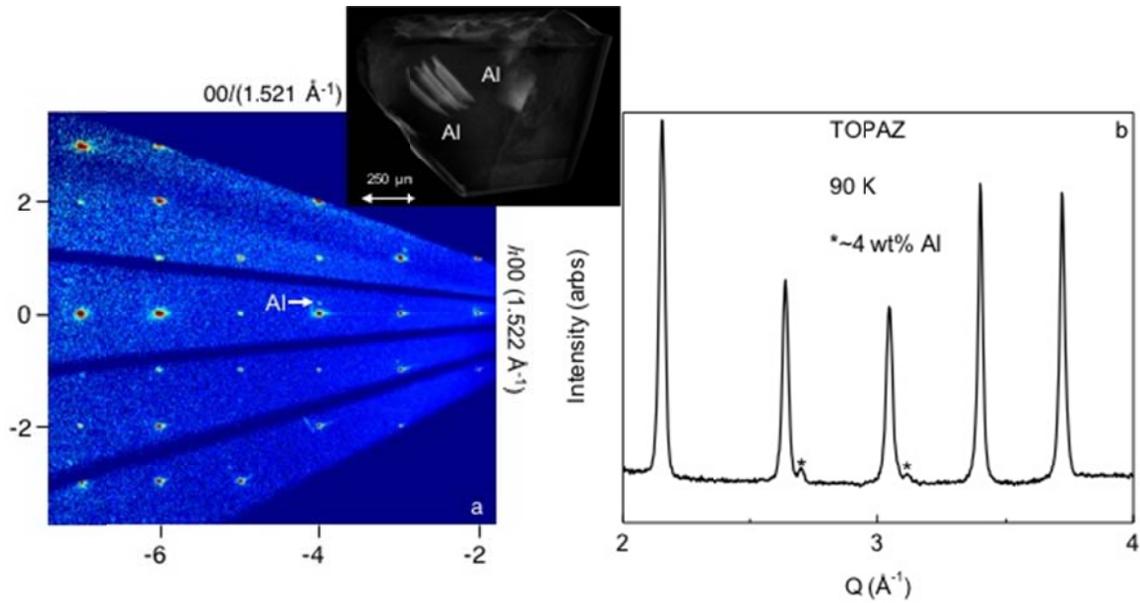

**Figure 2**. a) A 00*l* versus *h*00 precession image of the $^{154}$Sm$^{11}$B$_6$ flux grown crystal collected at *T* = 90 K using the TOPAZ single crystal diffractometer located at the Spallation Neutron Source and a X-ray CT image showing the presence of inclusions within the $^{154}$Sm$^{11}$B$_6$ crystal. The companion reflections (see white arrow) correspond to epitaxial aluminum present in this flux grown crystal, determined using b) a neutron diffraction histogram obtained from radial integration of the single crystal neutron diffraction data. The asterisks denote the reflections from the epitaxial aluminum present in the flux grown crystal of SmB$_6$.

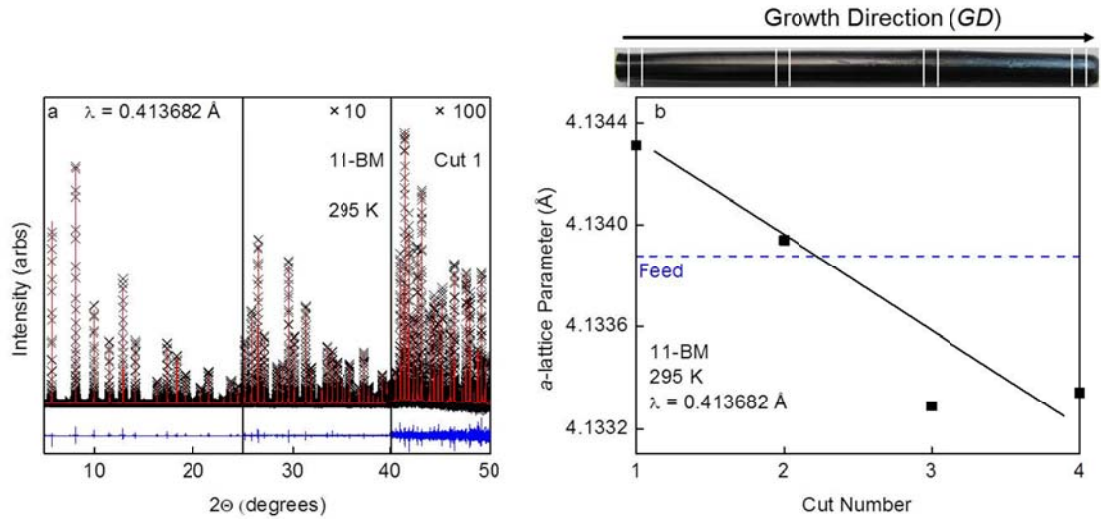

**Figure 3**. a) Rietveld refinement to synchrotron X-ray diffraction data at $T$ = 295 K collected on Cut 1 of the $SmB_6$ floating zone grown single crystal. The black crosses, red lines, and blue lines correspond to the collected data, refined model, and difference curve respectively. The higher angle data are multiplied by ×10 ($25 \geq 2\Theta \geq 40$) and ×100 ($40 \geq 2\Theta \geq 50$) to highlight the quality of the fit. Fits to cuts 2-4 are of similar quality. b) The refined $a$-lattice parameter values for cuts 1-4 versus cut number, where this value decreases with cut number. The line serves as a guide to the eye and the error bars are contained within the data points. Additionally, the refined $a$-lattice parameter value for the polycrystalline feed rod used for the growth of the *FZ* SmB6 crystal versus cut number (blue dashed line) is shown.

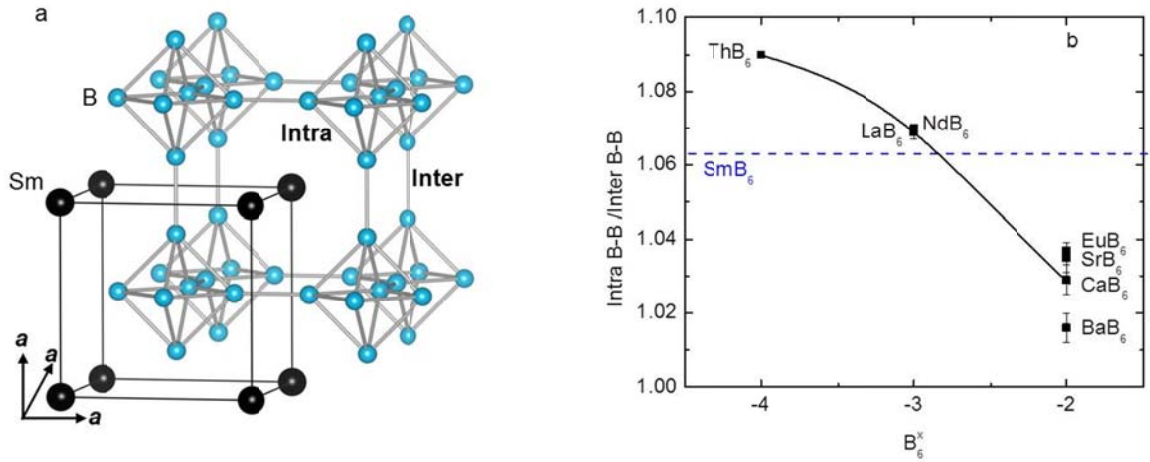

**Figure 4**. a) Structural depiction of $SmB_6$. b) The intra boron-boron bond distances normalized by the inter boron-boron distances (Intra B-B/Inter B-B) of the $B_6$ octahedral cluster for selected hexaborides versus different charges of the $B_6$ cluster ($B_6^x$) (■). Also shown is the Intra B-B/Inter B-B for all $SmB_6$ *FZ* cuts and the *FG* $SmB_6$ sample (blue dashed line). The charge on the $B_6$ cluster for $SmB_6$ mostly closely resembles -3, but is slight reduced, consistent with a partial mixed valency. The curved solid line serves as a guide to the eye.

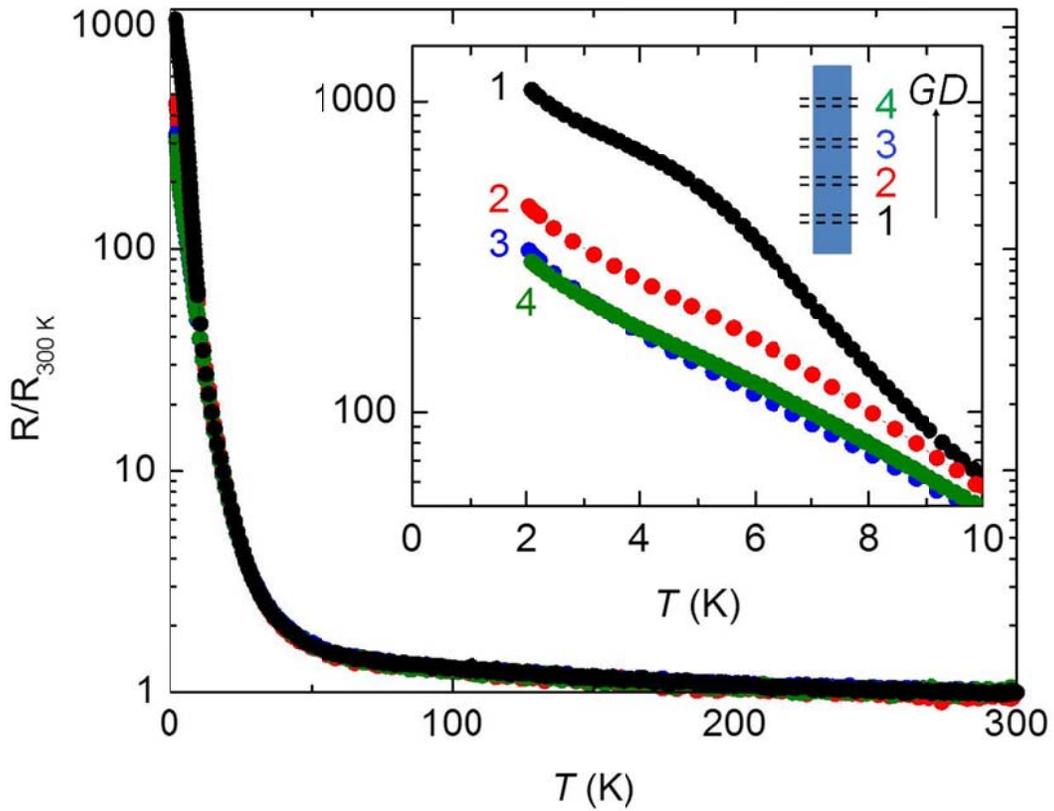

**Figure 5**. Resistance (R) normalized by the room temperature resistance values ($R_{300\,K}$) versus temperature (*T*) for cuts 1-4. The inset highlights the trend in normalized resistance from 0 to 10 K showing that the magnitude of $R/R_{300\,K}$ and the degree of plateauing decreases from 1 to 4.

**Supporting Information For: On the Chemistry and Physical Properties of Flux and Floating Zone Grown $SmB_6$ Single Crystals**

W. A. Phelan,[1,2,3*] S. M. Koohpayeh,[2] P. Cottingham,[1,2,3] J. A. Tutmaher,[1,2,3] J. C. Leiner,[4] M. D. Lumsden,[4] C. M. Lavelle,[5] X. P. Wang,[6] C. Hoffmann,[6] M. A. Siegler,[1] and T. M. McQueen[1,2,3*]

[1]*Department of Chemistry, Johns Hopkins University, Baltimore, MD 21218, USA*
[2]*Institute for Quantum Matter, Department of Physics and Astronomy, Johns Hopkins University, Baltimore, MD 21218, USA*
[3]*Department of Materials Science and Engineering, Johns Hopkins University, Baltimore, MD 21218, USA*
[4]*Quantum Condensed Matter Division, Oak Ridge National Laboratory, Oak Ridge, TN 37831, USA*
[5]*Applied Nuclear Physics Group, Johns Hopkins University, Applied Physics Laboratory, Laurel, MD,20723, USA*
[6]*Chemical and Engineering Materials Division, Neutron Sciences Directorate, Oak Ridge National Laboratory, Oak Ridge, TN 37831, USA*
[*]*wphelan2@pha.jhu.edu* and *mcqueen@jhu.edu*

**Figure S1.** A movie showing the reconstructed X-ray computed tomography (CT) frames of the $^{154}Sm^{11}B_6$ flux-grown crystal [photo only on arXiv]. It is obvious that this crystal is composed of two materials, one with a high atomic number (dark contrast) and low Z low atomic number (light contrast).

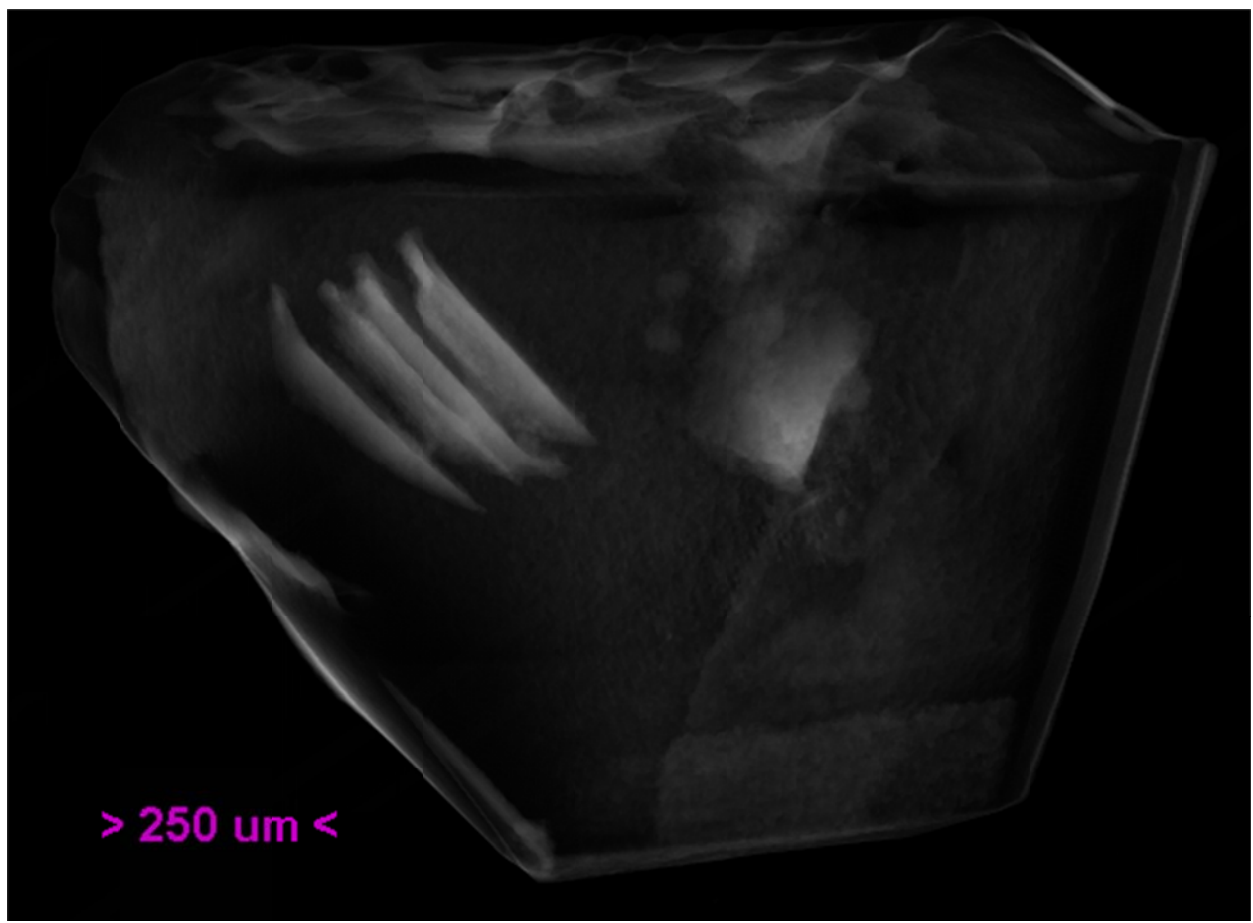

**Figure S2**. Rietveld refinements to synchrotron X-ray diffraction data at $T$ = 295 K collected on a) cut 2 b) cut 3, and c) cut 4 of the SmB$_6$ floating zone grown single crystal. The black crosses, red lines, and blue lines correspond to the collected data, refined model, and difference curve respectively. The higher angle data are multiplied by ×10 (25 ≥ 2Θ ≥ 40) and ×100 (40 ≥ 2Θ ≥ 50) to highlight the quality of the fit.

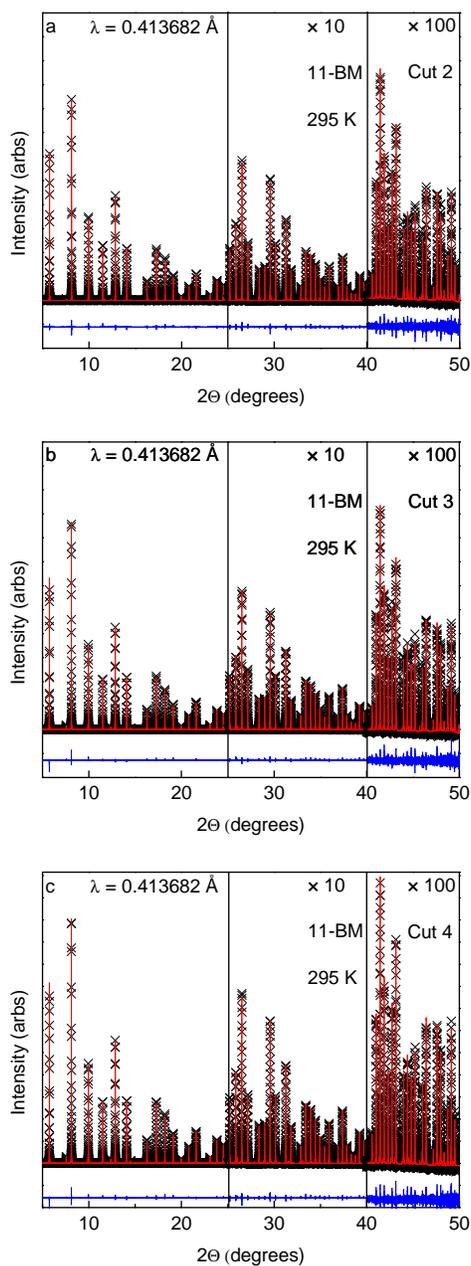

**Figure S3**. Rietveld refinements to X-ray diffraction data at $T$ = 295 K collected on the vaporized material, which amounted only to approximately 1% of the total material from the $SmB_6$ floating zone single crystal growth. The black crosses, red lines, and blue lines correspond to the collected data, refined model, and difference curve respectively. It is obvious from fits to this data that this vaporized material is a multi-phase mixture of $SmB_6$ (gray ticks) and $SmB_4$ (organic ticks).

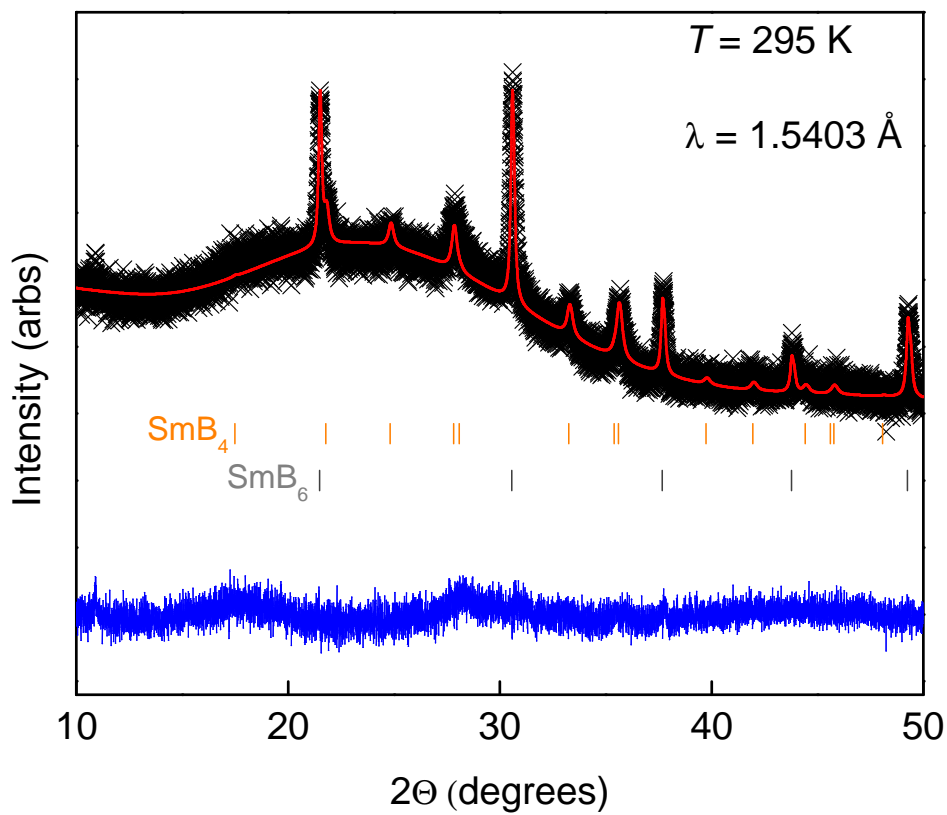

**Figure S4.** The (110), (310), and (640) reflections for cuts 1-4 where the peak positions were normalized along the x-axis relative to a silicon standard. All peak positions reside at higher angles when going from cuts 1 to 3 for each reflection, showing that the lattice parameters decrease with compositional variations along the crystal. The overlapping of the (110), (310), and (640) peak positions for cuts 3 and 4 show that these cuts have roughly the same lattice parameters and compositions.

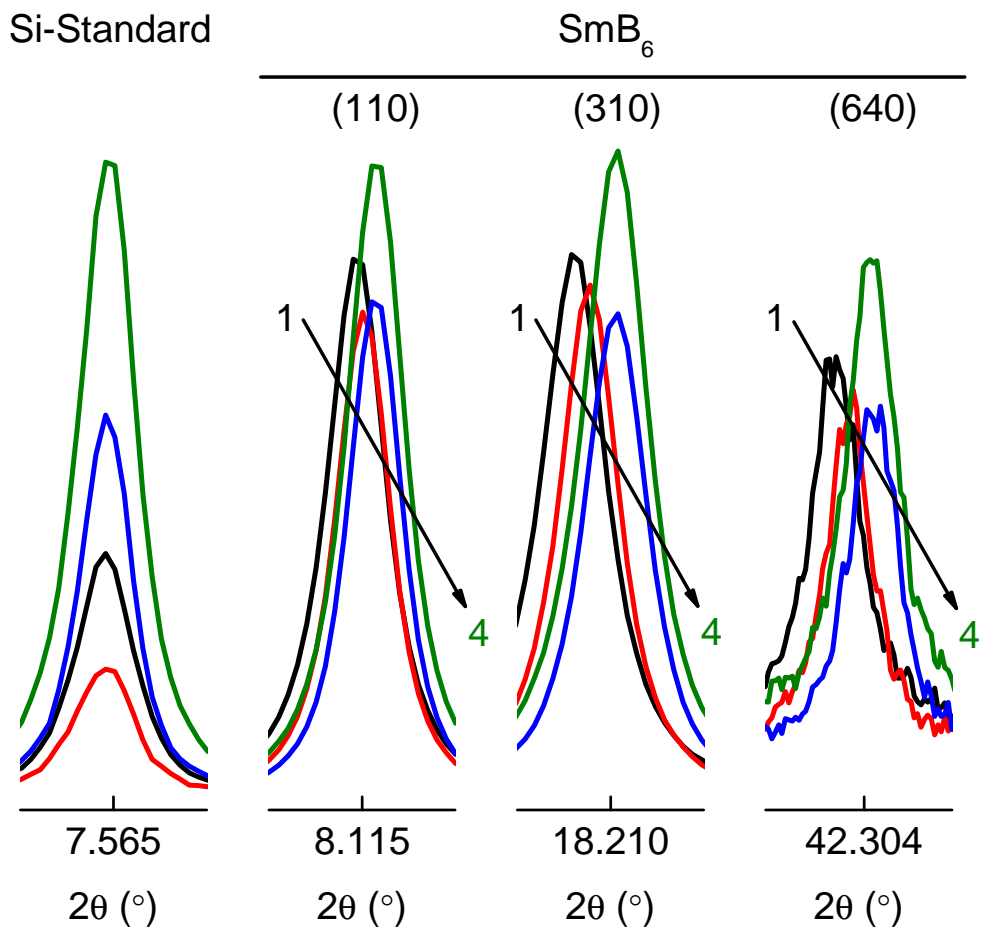

**Figure S5**. A plot of the resistance (R) normalized by the room temperature resistance values ($R_{300 K}$) versus temperature (*T*) for cuts 1'-3' (open circles) and cuts 1-3 (closed circles, Figure 5) from 0 to 10 K. To check the reproducibility of our resistance measurements, the authors removed the original platinum leads used to collect the data presented in Figure 5 for cuts 1-3, polished these three cuts, mounted new leads, and recollected the data (cuts 1'-3'). Additionally, resistances were measured using a new cut between the location of the original cut 1 and cut 2 (cut 2'') and a new cut beyond the location of the original cut 4 (cut 4''). Very much like the data for cuts 1-4 in Figure 5, the magnitude of the $R/R_{300 K}$ and the degree of plateauing decrease and the cut number gets larger for all cuts. Finally, the trend in lattice parameters of 4.1333(4) Å and 4.13284(3) Å for cut 2'' and cut 4'', respectively, agrees well with the resistance and lattice parameter trend for the original cuts 1-4.

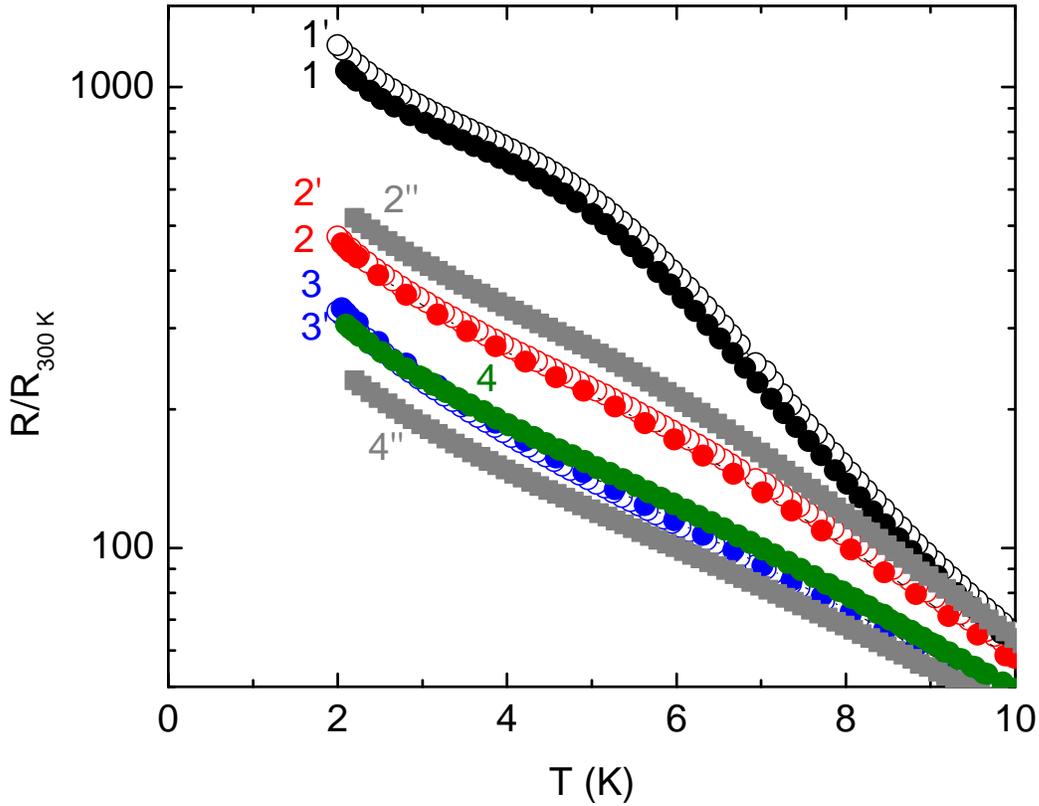

**Figure S6**. Concentration (ppm wt) of elements present in the starting material and cuts 1-3 versus Atomic Number. These semi-quantitative trace elemental analyses results were obtained from glow discharge mass spectrometry (GDMS) experiments and are tabulated in Table S5.

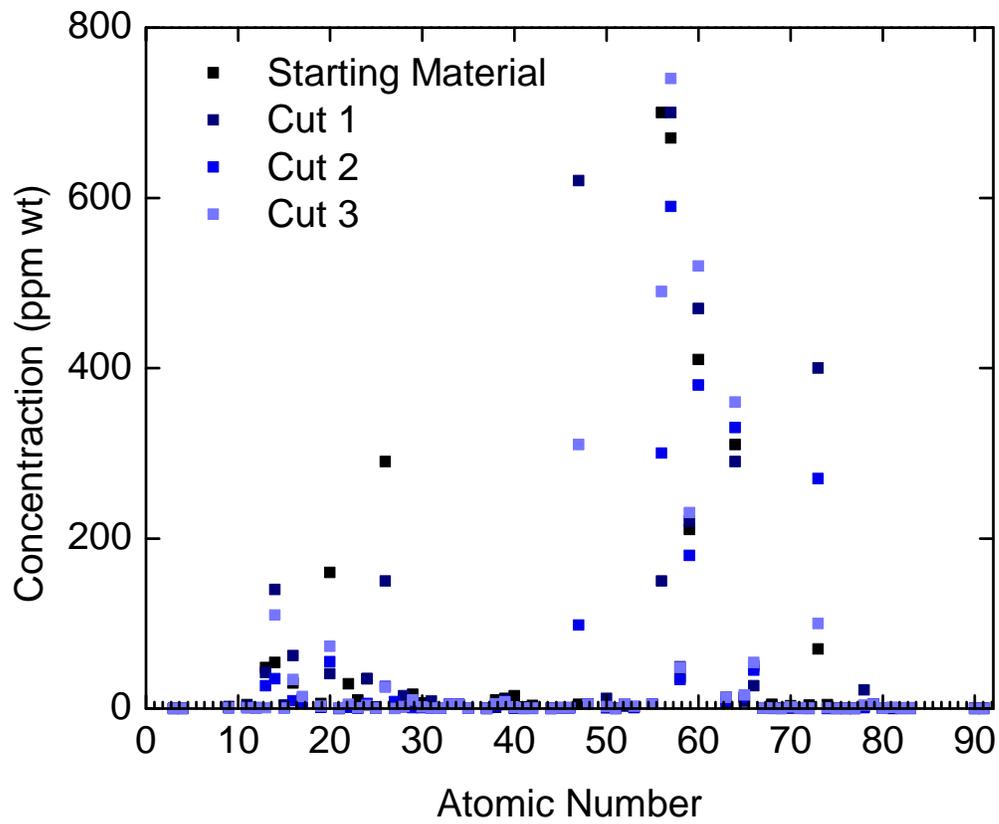

**Table S1**. Crystallographic parameters for the flux grown $^{154}$Sm$^{11}$B$_6$ crystal obtained from model fits to the single crystal X-ray diffraction data. The statistical uncertainties are given in parentheses.

| | |
|---|---|
| Temperature (K) | 293(2) |
| Composition | SmB$_{5.88}$ |
| Space group | *Pm-3m* |
| *a* (Å) | 4.13283(2) |
| *V* (Å$^3$) | 70.590(1) |
| *Z* | 1 |
| Collected Reflections | 7089 |
| Independent Reflections | 154 |
| *GooF* | 1.18 |
| $R_1[F^2 > 2\sigma(F^2)]$[a] | 0.009 |
| $wR_2(F^2)$[b] | 0.020 |
| $\Delta\rho_{max}$ (e Å$^{-3}$) | 1.00 |
| $\Delta\rho_{min}$ (e Å$^{-3}$) | -1.36 |

[a]$R_1(F) = \Sigma \,||F_o| - |F_c||/\Sigma\,|F_o|$; [b]$wR_2(F^2) = [\Sigma\,[w\,(F_o^2 - F_c^2)^2]/\Sigma\,[w\,(F_o^2)^2]]^{1/2}$

**Table S2.** Atomic fractional coordinates, site occupancies, and ADPs for flux grown $^{154}$Sm$^{11}$B$_6$ crystal obtained from model fits to the single crystal X-ray diffraction data. The statistical uncertainties are given in parentheses.

$T = 293(2)$ K

| atom | Wyckoff Site | x | y | z | Occupancy | $U_{11}$ (Å$^2$) | $U_{22}$ (Å$^2$) | $U_{33}$ (Å$^2$) |
|---|---|---|---|---|---|---|---|---|
| Sm1 | 1a | 0 | 0 | 0 | 1 | 0.00776(4) | 0.00776(4) | 0.00776(4) |
| B1 | 6f | 0.2000(1) | ½ | ½ | 0.98 | 0.0036(2) | 0.0036(2) | 0.00523(14) |

**Table S3.** The percent abundance for the differing isotopes of boron and samarium in the flux grown doubly enriched $^{154}$Sm$^{11}$B$_6$ crystal.

| | % Abundance of Boron | | % Abundance of Samarium | | | | | | |
|---|---|---|---|---|---|---|---|---|---|
| | $^{10}$B | $^{11}$B | $^{144}$Sm | $^{147}$Sm | $^{148}$Sm | $^{149}$Sm | $^{150}$Sm | $^{152}$Sm | $^{154}$Sm |
| Doubly Enriched SmB$_6$ Crystal | 3.6(1) | 96.4(2.1) | 0.0067(2) | 0.044(1) | 0.041(1) | 0.0550(3) | 0.0390(2) | 0.305(2) | 99.51(0.74) |

**Table S4.** The concentration of differing elements present in the SmB$_6$ starting material, cut 1, cut 2, and cut3.

| | Concentration (ppm wt) | | | |
|---|---|---|---|---|
| | Starting Material | Cut 1 | Cut 2 | Cut 3 |
| Li | <0.05 | <0.05 | <0.05 | <0.05 |
| Be | <0.05 | <0.05 | <0.05 | <0.05 |
| B | Bulk | Bulk | Bulk | Bulk |
| F | <1 | 2.2 | <1 | <1 |
| Na | 4.7 | 0.97 | 2.4 | 0.83 |
| Mg | 0.95 | 1.1 | 0.91 | 0.47 |
| Al | 48 | 42 | 27 | 0.99 |
| Si | 54 | 140 | 35 | 110 |
| P | 0.78 | 4.3 | 1.1 | 0.52 |
| S | 30 | 62 | 9.1 | 34 |
| Cl | ~10 | ~11 | ~7 | ~14 |
| K | 6.1 | 1.3 | 1.8 | 2.6 |
| Ca | 160 | 41 | 55 | 73 |
| Sc | 0.06 | 0.08 | <0.05 | 0.44 |
| Ti | 29 | 2.0 | 1.4 | 5.2 |
| V | 10 | 0.46 | 0.79 | 0.76 |
| Cr | 35 | 35 | 6.0 | 3.8 |
| Mn | 0.65 | 2.0 | 0.38 | 0.27 |
| Fe | 290 | 150 | 26 | 25 |
| Co | 0.64 | 2.4 | 8.2 | 0.19 |
| Ni | 13 | 15 | 1.9 | 2.1 |
| Cu | 17 | 4.3 | 0.81 | 9.8 |
| Zn | 6.5 | 0.72 | 1.9 | 1.8 |
| Ga | <0.5 | 8.5 | <0.5 | <0.5 |
| Ge | <0.5 | <0.5 | <0.5 | <0.5 |
| As | <5 | <5 | <5 | <5 |
| Se | <5 | <5 | <5 | <5 |
| Br | <0.5 | <0.5 | <0.5 | <0.5 |
| Rb | <0.05 | <0.05 | <0.05 | <0.05 |

**Table S4 continued.** The concentration of differing elements present in the SmB$_6$ starting material, cut 1, cut 2, and cut3.

| | Concentration (ppm wt) | | | |
|---|---|---|---|---|
| | Starting Material | Cut 1 | Cut 2 | Cut 3 |
| Sr | 10 | 1.5 | 3.8 | 5.5 |
| Y | 12 | 4.5 | 8.4 | 7.7 |
| Zr | 15 | 0.33 | 0.71 | 1.2 |
| Nb | 1.8 | 0.10 | <0.05 | 0.29 |
| Mo | 3.8 | 0.45 | <0.05 | 0.49 |
| Ru | <0.05 | 0.16 | <0.05 | <0.05 |
| Rh | <1 | <1 | <1 | <1 |
| Pd | <1 | <1 | <1 | <1 |
| Ag | 5.2 | 620 | 98 | 310 |
| Cd | <5 | <5 | <5 | <5 |
| In | Binder | Binder | Binder | Binder |
| Sn | <1 | 12 | <1 | 2.3 |
| Sb | <0.1 | <0.1 | <0.1 | <0.1 |
| Te | 4.2 | 2.7 | 4.8 | 5.4 |
| I | 1.8 | 0.73 | 1.2 | 2.9 |
| Cs | <5 | <5 | <5 | <5 |
| Ba | 700 | 150 | 300 | 490 |
| La | 670 | 700 | 590 | 740 |
| Ce | 37 | 49 | 34 | 48 |
| Pr | 210 | 220 | 180 | 230 |
| Nd | 410 | 470 | 380 | 520 |
| Sm | Bulk | Bulk | Bulk | Bulk |
| Eu | 13 | 6.6 | 9.5 | 13 |
| Gd | 310 | 290 | 330 | 360 |
| Tb | 13 | 8.7 | 13 | 16 |
| Dy | 50 | 27 | 45 | 54 |
| Ho | 0.48 | 0.30 | 1.1 | 0.44 |
| Er | 5.1 | 0.15 | 0.35 | 0.61 |
| Tm | 0.95 | <0.05 | <0.05 | 0.11 |

**Table S4 continued.** The concentration of differing elements present in the SmB$_6$ starting material, cut 1, cut 2, and cut3.

| | Concentration (ppm wt) | | | |
|---|---|---|---|---|
| | **Starting Material** | **Cut 1** | **Cut 2** | **Cut 3** |
| Yb | 3.2 | 0.35 | 1.5 | 1.7 |
| Lu | 0.94 | 0.08 | 0.25 | 0.51 |
| Hf | 4.1 | <0.1 | <0.1 | 0.63 |
| Ta | ≤70 | ≤400 | ≤270 | ≤100 |
| W | 4.7 | <0.1 | <0.1 | 1.0 |
| Re | <0.1 | <0.1 | <0.1 | <0.1 |
| Os | <0.1 | <0.1 | <0.1 | <0.1 |
| Ir | <0.1 | <0.1 | <0.1 | 0.19 |
| Pt | 1.8 | 22 | 1.1 | 3.9 |
| Au | <5 | <5 | <5 | <5 |
| Hg | <0.5 | <0.5 | <0.5 | <0.5 |
| Tl | <0.01 | 0.10 | 0.93 | 1.5 |
| Pb | <0.05 | 1.1 | 0.11 | 0.28 |
| Bi | <0.05 | 1.1 | 0.11 | 0.24 |
| Th | 0.04 | 0.03 | 0.04 | 0.07 |
| U | 0.03 | 0.01 | 0.02 | 0.03 |